\documentclass [prd,reprint,eqsecnum,nofootinbib]  {revtex4-1}
\usepackage{amsfonts,amsmath,amscd,mathrsfs,hyperref}

\begin{document}
\title{%
Local Maxwell symmetry and gravity}
\author{R. Durka}
\email{rdurka@ift.uni.wroc.pl}\affiliation{Institute for Theoretical
Physics, University of Wroc\l{}aw, Pl.\ Maxa Borna 9, Pl--50-204
Wroc\l{}aw, Poland}
\author{J. Kowalski-Glikman}
\email{jkowalskiglikman@ift.uni.wroc.pl}\affiliation{Institute for
Theoretical Physics, University of Wroc\l{}aw, Pl.\ Maxa Borna 9,
Pl--50-204 Wroc\l{}aw, Poland}
\date{\today}
\begin{abstract}\noindent
In this paper we discuss some physical aspects of a theory obtained
by gauging the AdS-Maxwell symmetry. Such theory has the form of
Einstein gravity coupled to the $\sf{SO(3,1)}$ Yang-Mills field. We
notice that there is another tetrad field, which can be associated
with  linear combination of Lorentz and Maxwell connections. Taking
this tetrad as a fundamental variable makes it possible to  cast the
theory into the form of $f-g$ gravity, in first order formulation.
Finally we discuss a simple cosmological model derived from the
AdS-Maxwell gravity.
\end{abstract}
\maketitle

\section{Introduction}

The Maxwell symmetry is an extension of Poincar\'e symmetry, being a
symmetry of fields on constant electromagnetic background
\cite{Bacry:1970ye}, \cite{Schrader:1972zd}. This symmetry is of
interest from purely algebraic point of view because it circumvents
a well known theorem that does not allow for central extension of
this algebra (see e.g., \cite{Galindo:1967}, \cite{Soroka:2011tc},
\cite{Gomis:2009dm}).

However, until recently  the Maxwell symmetry did not attract much
interest. This is somehow surprising, because physical systems
living on constant electromagnetic background, or closely to such a
situation related, are frequently encountered and important in
physics. This suggests that the Maxwell symmetry may play a quite
important role in physics. For example since the canonical
non-commutativity (for review see \cite{Douglas:2001ba}) is
ultimately related to the behavior of quantum systems on constant
electromagnetic background (or generalization thereof), one can
expect that this symmetry may play an important role in
non-commutative field theories (as already noticed in
\cite{Doplicher:1994tu}.) By the same token it does not seem
excluded that it might be relevant in the context of quantum Hall
effect.

Leaving these speculations to future works, in this paper we
investigate the related problem, namely the construction and
properties of theories with local Maxwell invariance. In the next
section, following \cite{Durka:2011nf}, we recall the formulation of
such a theory, deriving also its field equations. The following
short section is devoted to some comments on possible relations
between the gauged Maxwell theories and bi-metric theories of
gravity. Finally, in section IV we discuss a simple class of
cosmological solutions of this theory.

\section{Gauged Maxwell theory}

We begin this section presenting some definitions and then recalling
some of the results presented in \cite{Durka:2011nf}.

The starting point of our construction is a generalization of
Maxwell symmetry, in which the Poincar\'e subalgebra of the original
Maxwell algebra \cite{Bacry:1970ye}, \cite{Schrader:1972zd} is
replaced with the anti de Sitter counterpart. In other words we
consider the algebra of symmetries of (complex scalar, Dirac) fields
on anti de Sitter space and constant electromagnetic
background\footnote{A detailed discussion will be presented in a
forthcoming paper.}. The algebra of such symmetries, which turns out
to be  isomorphic to $\sf{so}(3,1)\oplus\sf{so}(3,2)$ was discussed
previously in \cite{Soroka:2006aj}, \cite{Bonanos:2010fw},
\cite{Lukierski:2010dy} and has the following form
\cite{Durka:2011nf}
\begin{align}
&[\mathcal{P}_{a},\mathcal{P}_{b}]=i(\mathcal{M}_{ab}-\mathcal{Z}_{ab})\,
,\nonumber\\
&[\mathcal{M}_{ab},\mathcal{M}_{cd}]=-i(\eta_{ac}\mathcal{M}_{bd}+\ldots)\, ,\nonumber\\
&[\mathcal{M}_{ab},\mathcal{Z}_{cd}]=-i(\eta_{ac}\mathcal{Z}_{bd}+\ldots)\,,\nonumber\\
&[\mathcal{Z}_{ab},\mathcal{Z}_{cd}]=-i(\eta_{ac}\mathcal{Z}_{bd}+\ldots)\, ,\nonumber\\
&[\mathcal{M}_{ab},\mathcal{P}_c]=-i(\eta_{ac}\mathcal{P}_{b}-\eta_{bc}\mathcal{P}_{a}),\quad
[\mathcal{Z}_{ab},\mathcal{P}_{c}]=0\,,\label{1}
\end{align}
where $\ldots$ denotes three more terms obtained by
antisymmetrization in the pairs of indices $(ab)$ and $(cd)$. The
original Maxwell algebra \cite{Soroka:2006aj},
\cite{Bonanos:2010fw}, \cite{Lukierski:2010dy} can be easily
obtained from (\ref{1}) by contraction, which makes $\mathcal{M}$ on
the right hand side of the first commutator and the whole right hand
side of the fourth one vanish.

This algebra can be gauged by defining the gauge field (connection)
\begin{equation}\label{2}
\mathbb{A}_{\mu}=\frac{1}{2}\omega_\mu{}^{ab}
\mathcal{M}_{ab}+\frac{1}{\ell}e_\mu{}^{a}
\mathcal{P}_{a}+\frac{1}{2}h_\mu{}^{ab} \mathcal{Z}_{ab}\,,
\end{equation}
and its curvature
\begin{equation}\label{3}
\mathbb{F}_{\mu\nu}=\partial_\mu\mathbb{A}_{\nu}-\partial_\nu\mathbb{A}_{\mu}-i[\mathbb{A}_{\mu},\mathbb{A}_{\nu}]\,,
\end{equation}
which can be decomposed into Lorentz, translational, and Maxwell
parts
\begin{equation}\label{4}
\mathbb{F}_{\mu\nu}=\frac12\, F_{\mu\nu}^{ab}\, \mathcal{M}_{ab}
+\frac{1}{\ell}T_{\mu\nu}^{a}\,\mathcal{P}_{a}+ \frac12\,
G_{\mu\nu}^{ab}\, \mathcal{Z}_{ab}\,,
\end{equation}
where
\begin{eqnarray}
F_{\mu\nu}^{ab}&=&R^{ab}_{\mu\nu}+\frac{1}{\ell^2}( e^a_\mu e^b_\nu- e^a_\nu e^b_\mu) \,,\label{5a}\\
T_{\mu\nu}^{a}&=&D^\omega_\mu e^a_\nu -D^\omega_\nu e^a_\mu\,,\label{5b}\\
G_{\mu\nu}^{ab}&=&D^\omega_\mu h^{ab}_\nu -D^\omega_\nu
h^{ab}_\mu\nonumber-\frac{1}{\ell^2}( e^a_\mu e^b_\nu- e^a_\nu
e^b_\mu)\\ &+&(h^{ac}_\mu h^{\quad b}_{\nu\,c}-h^{ac}_\nu h^{\quad
b}_{\mu\,c})\, .\label{5c}
\end{eqnarray}
In the formula above we denote by $$D^\omega_\mu(\ast)\equiv
\partial_\mu(\ast)-i[\omega_\mu,(\ast)]$$ the covariant
derivative of the Lorentz connection $\omega$.

The gauge transformations of the gauge fields $h_\mu{}^{ab}$,
$\omega_\mu{}^{ab}$, $e_\mu{}^{a}$ can be calculated to be
\begin{eqnarray}
        \delta_\Theta h^{ab}_{\mu}&=&D^\omega_\mu \tau^{ab}+
h^{ac}_\mu(\lambda+\tau)_{c}^{\;\,b}+h^{bc}_\mu(\lambda+\tau)^{a}_{\;\;c}\label{6}\quad\\
      \delta_\Theta \omega^{ab}_{\mu}&=& D^\omega_\mu \lambda^{ab}
      \label{7}\\
   \delta_\Theta e^{a}_{\mu}&=&-\lambda^a_{\;b}
    \,e^b_\mu\,,\label{8}
\end{eqnarray}
where $\lambda^{ab}$ and $\tau^{ab}$ denote the parameter of local
 Lorentz and Maxwell symmetries, respectively. The
components of the curvature transform in a homogeneous way
\begin{eqnarray}
        \delta_\Theta G^{ab}_{\mu\nu}&=&-[\tau,F_{\mu\nu}]^{ab}-[(\lambda+\tau),G_{\mu\nu}]^{ab}\label{9}\\
      \delta_\Theta F^{ab}_{\mu\nu}&=&-[\lambda,F_{\mu\nu}]^{ab}  \label{10}\\
    \delta_\Theta
    T^{a}_{\mu\nu}&=&-\lambda^a_{\;b}T^b_{\mu\nu}\,.\label{11}
\end{eqnarray}
In the formulas above we omitted local translations, which, as
usual, are being eventually traded for general coordinate invariance
and are not symmetries of the action.

It is clear from eqs.\ (\ref{8}), (\ref{9})--(\ref{11}) that the
only geometrical action i.e., a Lorentz and Maxwell invariant
four-form that can be built from gauge fields and curvatures (see
\cite{Durka:2011nf} and references therein)  has the form
\begin{equation}
S_{E} =\frac1{64\pi G}\,
\int\epsilon_{abcd}\left(R_{\mu\nu}{}^{ab}e_{\rho}^{c}e_{\sigma}^{d}
-\frac{\Lambda}{3}e_{\mu}^{a}e_{\nu}^{b}e_{\rho}^{ c}e_{\sigma}^{
d}\right)\epsilon^{\mu\nu\rho\sigma}\label{12}
\end{equation}
being the standard Einstein action with a cosmological term. Of
course, there are other possible invariant terms, which are
quadratic in the curvatures; these terms are however topological
 and do not influence equations of motion.

Let us comment at this point the relation between our construction
and that presented in the recent paper \cite{deAzcarraga:2010sw}. In
that paper the authors construct the curvatures similar to the ones
we presented above (\ref{5a})--(\ref{5c}), but then, to built the
action, they use all possible geometric combinations that are local
Lorentz invariant. In this way the theory considered in
\cite{deAzcarraga:2010sw}
 breaks the Maxwell symmetry manifestly.
For this reason the geometric action derived in
\cite{deAzcarraga:2010sw}  is much reacher than the one derived in
\cite{Durka:2011nf} and presented above.

The action (\ref{12}) does not include curvature of the Maxwell
field $G$ (as said above all the geometrical terms including $G$ are
the topological ones) and one wonders if one could find nontrivial
terms which include the Maxwell field strength. Inspecting
(\ref{9}), (\ref{10}) it is easy to see that there are only two
possible terms of this sort. The first is the Yang-Mills action with
the $\sf{SO(3,1)}$ gauge group, already found in
\cite{Soroka:2011tc}, \cite{Durka:2011nf}
\begin{equation}\label{13}
 S_{1}= -  \frac1{4\varrho_1}\,\int d^4x\, e\, \big( F_{\mu\nu}{}^{ab} + G_{\mu\nu}{}^{ab}\big) \big( F^{\mu\nu}{}_{ab} +
    G^{\mu\nu}{}_{ab}\big)\, .
\end{equation}
Noting that $\epsilon^{abcd}$ is an  invariant tensor of
$\sf{SO(3,1)}$, we can construct yet another manifestly gauge
invariant action
\begin{equation}\label{13a}
 S_{2}= -  \frac1{4\varrho_2}\,\int d^4x\, e\, \big( F_{\mu\nu}{}^{ab} + G_{\mu\nu}{}^{ab}\big) \big( F^{\mu\nu}{}^{cd} +
    G^{\mu\nu}{}^{cd}\big) \epsilon_{abcd}\, .
\end{equation}
In the formulas above $\varrho_1$ and $\varrho_2$ are two
dimensionless coupling constants.

Using the explicit form of the curvatures $F$ and $G$, (\ref{5a})
and (\ref{5c}), the form of these actions can be further simplified.
Indeed
$$
F_{\mu\nu}{}^{ab}(\omega, e) + G_{\mu\nu}{}^{ab}(h,\omega,e) =
H_{\mu\nu}{}^{ab}(\varpi)
$$
where $H_{\mu\nu}{}^{ab}(\varpi)$ is the curvature tensor of the
connection $\varpi_{\mu}{}^{ab}\equiv
\omega_{\mu}{}^{ab}+h_{\mu}{}^{ab}$. From now on we consider
$\varpi$ to be the new fundamental variable. The total action is the
sum of the Einstein one (\ref{12}) and the actions for the field
$\varpi$
\begin{align}
S &= S_E + S_1 +S_2 \nonumber\\
&=\frac{1}{64\pi
G}\int\epsilon_{abcd}\left(R_{\mu\nu}{}^{ab}e_{\rho}^{c}e_{\sigma}^{d}
-\frac{\Lambda}{3}e_{\mu}^{a}e_{\nu}^{b}e_{\rho}^{ c}e_{\sigma}^{
d}\right)\epsilon^{\mu\nu\rho\sigma}\nonumber\\
&-\frac14\,\int  e\, \left(\frac1{\varrho_1}\, H_{\mu\nu}{}^{ab}
H^{\mu\nu}{}_{ab}+ \frac1{\varrho_2}\,H_{\mu\nu}{}^{ab}
H^{\mu\nu}{}^{cd}\,\epsilon_{abcd}\right)\, .\label{14}
\end{align}

The field equations resulting from this action are Einstein
equations (with cosmological constant term)
\begin{equation}\label{15a}
    R_{\mu\nu}-\frac12\, g_{\mu\nu}\, R +\Lambda\, g_{\mu\nu}= 8\pi G T_{\mu\nu}
\end{equation}
whose right hand side is the energy momentum tensor for the field
$H$
\begin{align}
    T_{\mu\nu} &= \frac{1}{\varrho_1}\left(H_{\mu\lambda}{}^{ab} H_\nu{}^\lambda{}_{ab} -
    \frac14\, g_{\mu\nu}\,H_{\lambda\sigma}{}^{ab}
    H^{\lambda\sigma}{}_{ab}\right) \nonumber\\
    &+\frac{1}{\varrho_2}\left(H_{\mu\lambda}{}^{ab} H_\nu{}^\lambda{}^{cd} -
    \frac14\, g_{\mu\nu}\,H_{\lambda\sigma}{}^{ab}
    H^{\lambda\sigma}{}^{cd}\right)\epsilon_{abcd}\label{15}
\end{align}

The field equations for $\varpi$ differ slightly from the standard
Yang-Mills form and read
\begin{equation}\label{16}
    \nabla^\mu \left(\frac1{\varrho_1}\, H_{\mu\nu}{}^{ab} +\frac1{\varrho_2}\,\epsilon^{abcd}
    H_{\mu\nu}{}_{cd}\right)=0\,,
\end{equation}
where $\nabla$ is the covariant derivative of the metric constructed
from the tetrad $e^a_\mu$ (as a result of field equations for the
action (\ref{14}) the spacetime torsion vanishes) acting on
spacetime indices, and of connection $\varpi$ acting on the algebra
indices.

Unfortunately the gauge theory with a non-compact gauge group is
known to suffer from the unitarity problem at the quantum level
\cite{Hsu:1980zt}. Moreover, already in the classical case their
coupling to gravity is problematic because the energy density is not
positive definite, as can be seen explicitly by inspecting the
$T_{00}$ component of the energy-momentum tensor in (\ref{15})
above. We will discuss this in more details in Sect.\ IV while
investigating the properties of simple cosmological solutions of the
theory defined by the action (\ref{14}).

\section{Bi-metric interpretation}

In this section we briefly discuss the relation between the gauged
Maxwell theory and bi-metric theories. To begin with let us notice
that the gauge transformations of the connection $\varpi$ introduced
in the preceding section have, formally, exactly the same form as
the ones of the gravitational connection $\omega$. Therefore we can
associate with it another ``tetrad'' field one form $f^a$ and the
``torsion'' two-form $\tau^a$
\begin{equation}\label{17}
    df^a + \varpi^a{}_b\wedge f^b=\tau^a\,.
\end{equation}
Having defined the new tetrad field $f^a$ it is possible to
construct yet another gauge invariant action, which is exactly the
Einstein action, but this time for the fields $f$ and $\varpi$.
Indeed
\begin{equation}
S_{E'} =\frac1{64\pi G'}\,
\int\epsilon_{abcd}\left(H_{\mu\nu}{}^{ab}f_{\rho}^{c}f_{\sigma}^{d}
-\frac{\Lambda'}{3}f_{\mu}^{a}f_{\nu}^{b}f_{\rho}^{ c}f_{\sigma}^{
d}\right)\epsilon^{\mu\nu\rho\sigma}\label{18}
\end{equation}
is manifestly gauge invariant, exactly as the Einstein action
(\ref{12}) is. The theory defined  by the sum of the actions $S_E$
and $S_{E'}$ reminds very much the $f-g$ theory of Isham, Salam, and
Strathdee \cite{Isham:1971gm} (for recent discussion and references
see e.g., \cite{Banados:2008fi}) in the first order formalism. The
only part that is missing is the contact term between tetrads $e$
and $f$. However, since the tetrads $e$ and $f$ transform under
different gauge groups it is only possible to couple the metrics
constructed out of them $g_{\mu\nu} = e_\mu{}^ae_\nu{}^b\eta_{ab}$
and $q_{\mu\nu} = f_\mu{}^af_\nu{}^b\eta_{ab}$. In the usual $f-g$
theory setup such constant terms break the $(diff)^2$ invariance of
the action $S_E+S_{E'}$ down to its diagonal subgroup. This is
however not the case here: our theory from the very start has just
one built-in diffeomorphism invariance that results from the gauged
translations. Therefore adding the contact term does not lead to
breaking of any symmetries of our theory. The most general contact
term has the form
\begin{equation}\label{19}
    S_{cont} =
   {M^2}(-g)^u(-q)^{(1/2-u)}\, V(g^{-1}q)
\end{equation}
where $(g^{-1}q)^\mu{}_\nu = g^{\mu\rho}q_{\rho\nu}$ and $V$ is an
arbitrary scalar potential that can be constructed from this tensor.

The $f-g$ theories have been investigated recently from many
perspectives, e.g., in \cite{Banados:2008fi}, \cite{Damour:2002wu},
\cite{Damour:2002ws}, \cite{ArkaniHamed:2002sp}, \cite{Blas:2005yk},
\cite{Speziale:2010cf}; see also \cite{Clifton:2011jh} for a recent
comprehensive review. We leave it to the future work to check if in
any of these perspective (brane worlds, Kaluza-Klein models,
noncommutative geometry, \ldots) the local Maxwell symmetry can find
its natural habitat.

\section{Example: a simple cosmological model}

In this section we  consider, as an example, a simple cosmological
solution of the model described by the action (\ref{14}). In the
course of this exercise we will encounter the problem of
non-positively defined energy, mentioned in Sect.\ II.

We make use of the simplest, flat FRW metric, which has the form
\begin{equation}\label{20}
    ds^2 =-N(t)^2dt^2+a(t)^2 (dx^2+dy^2+dz^2)
\end{equation}
In what follows we  analyze the dynamics of this model using as a
starting point the variational principle, not the equations of
motion, and therefore we substitute the metric (\ref{20}) into the
Einstein action (\ref{12}) and neglecting the (infinite) space
volume prefactor we obtain
\begin{equation}
S_E=\frac{1}{8\pi G}\int dt \left(-\frac{3}{N}a\dot a^2-\Lambda a^3
N\right)\,.\label{21}
\end{equation}

Let us now turn to the gauge fields part of the action. The first
step is to find the most general form of the fields, consistent with
the background symmetries of the FRW spacetime (\ref{20}). The
desired form is easy to guess. Indeed the only objects that could be
present are tensors invariant under action of the group of space
symmetries, which in this case is just the Euclidean group.
Therefore we take the only non-vanishing components of the
connection $\varpi$ to be
\begin{equation}\label{22}
    \varpi_i{}^{0\alpha} = f(t)\, \delta_i^\alpha\,,
    \quad \varpi_i{}^{\alpha\beta} = g(t)\,
    \epsilon_i{}^{\alpha\beta}\,,\quad \alpha,\beta=1,2,3\,.
\end{equation}
In the Appendix we prove that this ansatz is indeed correct.

Substituting this and the form of the metric (\ref{20}) to the
actions $S_1$ in (\ref{13}), we get
\begin{equation}
    S_1 =\frac1{\varrho_1}\int dt \left( \frac{3a}{N}\Big(\dot{g}^2 -\dot{f}^2\Big)
+ \frac{3N}{a} g^2\Big(4f^2-g^2\Big) \right)\label{23}
\end{equation}
Similarly, the action $S_2$ in (\ref{14}) takes the form
\begin{equation}
    S_2 =\frac1{\varrho_2}\int dt \left( \frac{12a}{N}\, \dot f\,\dot g - \frac{24N}{a}\, f\, g^3\right)\label{24}
\end{equation}
We see that in the action (\ref{23}) the sign of the  kinetic term
is wrong and the potential is not bounded from below, which, as
discussed above, exhibit the problems arising in the case of a
non-compact gauge group.

The Einstein equations can be obtained from the action $S_E
+S_1+S_2$ by varying over the scale factor $a(t)$ and the lapse
function $N(t)$. Since the latter serves as a Lagrange multiplier,
in the resulting equation one can set $N(t)=1$. As a result we get
\begin{align}
\left(\frac{\dot{a}}{a} \right)^2-\frac{\Lambda}{3}
 &=\frac{8\pi G}{3\varrho_1}\left[\frac{3}{a^2} \left(\dot g^2 - \dot
 f^2\right)
 -\frac{3}{a^4}\,g^2\left(4f^2-g^2\right)\right]\nonumber\\
 &+\frac{8\pi G}{3\varrho_2}\left[\frac{12}{a^2}\, \dot f \dot g +\frac{24}{a^4}\,f\, g^3\right]\,.\label{25}
\end{align}
and
\begin{align}
\frac{2\ddot a}{a} &+ \left(\frac{\dot a}{a}\right)^2 - \Lambda
 =-\frac{8\pi G}{3\varrho_1}\left[\frac{3}{a^2} \left(\dot g^2 - \dot
 f^2\right)\right.\nonumber\\
 &-\left.\frac{3}{a^4}\,g^2\left(4f^2-g^2\right)\right]
 -\frac{8\pi G}{3\varrho_2}\left[\frac{12}{a^2}\, \dot f \dot g +\frac{24}{a^4}\,f\, g^3\right]\label{26}
\end{align}
Comparing these equations with the Friedmann equations in the
canonical form
$$
\left(\frac{\dot{a}}{a} \right)^2-\frac\Lambda3=\frac{8\pi
G}{3}\varepsilon\, , \quad \ddot a -\frac{\Lambda a}3=-\frac{4\pi
G}{3}(\varepsilon+3p)a
$$
we see that in our case the energy density is
\begin{align}
    \varepsilon&= \frac{1}{\varrho_1}\left[\frac{3}{a^2} \left(\dot g^2 - \dot
 f^2\right)
 -\frac{3}{a^4}\,g^2\left(4f^2-g^2\right)\right]\nonumber\\
 &+\frac{1}{\varrho_2}\left[\frac{12}{a^2}\, \dot f \dot g +\frac{24}{a^4}\,f\, g^3\right]\label{27}
\end{align}
and clearly is not positive definite. As for the pressure,
\begin{equation}\label{28}
    p=\frac\varepsilon3
\end{equation}
as it should be, because the field $\varpi$ is a massless Yang-Mills
field. Therefore eqs.\ (\ref{25}), (\ref{26}) describe a universe
with cosmological constant filled with radiation. Therefore the
solutions are
\begin{align*}
 &a(t)= C \sqrt{2 (t-t_0)}\,, &\Lambda=0\\
 &a(t)= C \sqrt{\cos \left(\sqrt{\frac{|\Lambda|}{3} } \left(2 (t-t_0)\right)\right)}\,, &\Lambda<0\\
&a(t)= C \sqrt{\cosh \left(\sqrt{\frac{|\Lambda|}{3} } \left(2
(t-t_0)\right)\right)}\,, &\Lambda>0\,.
\end{align*}
Notice that the form of the first Friedmann equation (\ref{25})
forces the energy density to be strictly positive or positive
definite (for negative and vanishing $\Lambda$) or bounded from
below (for positive $\Lambda$). Therefore the coupling to the
gravitational field seems to exclude some possible configurations of
the field $\varpi$ with large negative energy density. It would be
interesting to see if this is a consequence of the cosmological
ansatz (\ref{20}), or a somehow generic result.

For completeness we write the $g$ and $f$ equations of motion. They
read
\begin{equation}
\frac1{\varrho_1}\left[ -(a\dot g)\dot{} + \frac{4}a\, gf^2 -
\frac{2}a\, g^3\right] -\frac1{\varrho_2}\left[ (2a\dot f)\dot{} +
\frac{12}a\, fg^2 \right]=0\label{29}
\end{equation}
and
\begin{equation}
\frac1{\varrho_1}\left[ (a\dot f)\dot{} + \frac{4}a\, g^2f \right]
-\frac1{\varrho_2}\left[ (2a\dot g)\dot{} + \frac{4}a\, g^3
\right]=0\label{30}
\end{equation}
It is worth noticing that these equations are related to the
continuity equation
\begin{equation}\label{31}
\dot \varepsilon +3H(\varepsilon+p)= \dot \varepsilon
+4H\varepsilon=0\,,
\end{equation}
where we used (\ref{28}). Indeed substituting (\ref{27}) to
(\ref{31}) one finds
\begin{equation}\label{32}
    \dot f\, \mbox{\framebox{(\ref{29})}} +\dot g\, \mbox{\framebox{(\ref{30})}}=0\,.
\end{equation}

For the simplest possible solution of these equations, with $f$ and
$g$ being time-independent constants,  excluding the trivial
possibility $g=0$, we find from (\ref{30}) that
$$f=\frac{\varrho_1}{\varrho_2}\,g\,.$$ Eq.\ (\ref{29}) tells
that a nontrivial solution is possible only
if$$\left(\frac{\varrho_1}{\varrho_2}\right)^2=-\frac1{4}\,,$$ so
that the solution does not exist. Similar conclusion can be reached
if one takes $\dot g,\, \dot f \sim a^{-1}$. This suggests that the
dynamics of the fields $f$ and $g$ is quite nontrivial and will be
discussed elsewhere.

\appendix*
\section{Cosmological gauge field}

In this appendix we derive the most general form of the gauge field
$h_\mu^{ab}$ on the flat FRW background. The infinitesimal
symmetries of this background are spacial translations and
rotations, forming together the three dimensional Euclidean group
with the algebra (since the metric on the constant time section of
the flat FRW spacetime is, up to the rescaling, just $\delta_{ij}$,
we will be not very careful in placing indices up or down)
\begin{equation}\label{a1}
    [R_i, R_j] = \epsilon_{ijk}\, R_k, \, [R_i, T_j] = \epsilon_{ijk}\, T_k, \, [T_i, T_j]
    =0\, .
\end{equation}
The Killing vector fields of the flat FRW spacetime form a
representation of this algebra
\begin{align}
T_i &= \frac{\partial}{\partial x^i}\nonumber \\ R_i
&=\epsilon_{ijk}\, x^j \,\frac{\partial}{\partial x^k}\label{a2}
\end{align}

The gauge field invariance under the action of symmetries of
spacetime means that the Lie derivatives of the field along the
Killing vectors of the symmetries are zero, up to the gauge
transformations, i.e. (for general theory see \cite{Forgacs:1979zs}
and \cite{Jackiw:1979ub})
\begin{align}
    {\cal L}_{T_i}\,h_\mu^{ab} &= \partial_\mu \lambda^{(T)}_i{}^{ab}
    + [h_\mu,\lambda^{(T)}_i]^{ab}\,,\label{a3}\\  {\cal L}_{R_i}\,h_\mu^{ab} &= \partial_\mu \lambda^{(R)}_i{}^{ab}
    + [h_\mu,\lambda^{(R)}_i]^{ab}\,.\label{a4}
\end{align}
There are important integrability conditions for these equations
resulting from the identity
$$
[{\cal L}_X, {\cal L}_Y] = {\cal L}_{[X,Y]}\, ,
$$
which leads to the following condition on the gauge parameters
$\lambda$
\begin{equation}\label{a5}
{\cal L}_X\lambda^{(Y)}-{\cal L}_Y\lambda^{(X)}
-[\lambda^{(X)},\lambda^{(Y)}] =\lambda^{([X,Y])}\,.
\end{equation}
In the case of two translations equation (\ref{a5}) takes the form
\begin{equation}\label{a6}
    \partial_i\lambda^{(T)}_j{}^{ab}-
    \partial_j\lambda^{(T)}_i{}^{ab} -
    [\lambda^{(T)}_i,\lambda^{(T)}_j]^{ab}=0\, .
\end{equation}
This equation tells that the ``field strengths'' of the ``gauge
field'' $\lambda^{(T)}_i{}^{ab}$ vanishes, and since the equation
(\ref{a3}) is gauge-covariant, it follows that
$\lambda^{(T)}_j{}^{ab}$ must vanish itself.

Then the equation (\ref{a5}) applied to rotations and translations
reduces to the condition
$$
\partial_i\lambda^{(R)}_j{}^{ab}=0
$$
and therefore $\lambda^{(R)}_j{}^{ab}$ is just a constant. Then the
equation (\ref{a5}) applied to rotations takes the form
\begin{equation}\label{a7}
    [\lambda^{(R)}_i,\lambda^{(R)}_j]^{ab}=\epsilon_{ijk}\,
    \lambda^{(R)}_k{}^{ab}\,,
\end{equation}
which means that the matrices $\lambda^{(R)}_i{}^{ab}$ form a
representation of the rotation group $\sf{SO(3)}$.

Knowing this we can now return to the equations (\ref{a3}),
(\ref{a4}) which now take a much simpler form
\begin{align}
    {\cal L}_{T_i}\,h_\mu^{ab} &= 0\label{a8}\\  {\cal L}_{R_i}\,h_\mu^{ab} &=  [h_\mu,\lambda^{(R)}_i]^{ab}\label{a9}
\end{align}
Equation (\ref{a8}) tells that the gauge field depends only on time
$h_\mu^{ab}(x,t) = h_\mu^{ab}(t)$. Then from (\ref{a9}) it follows
that $h_0^{ab}$ commutes with all the matrices $\lambda^{(R)}_i$ and
therefore must be zero, and
\begin{equation}\label{a10}
\epsilon_{ijk}\, h_j^{ab}=[h_k,\lambda^{(R)}_i]^{ab}
\end{equation}
Returning to eq.\ (\ref{a7}) we see that the matrices
$\lambda^{(R)}_i{}^{ab}$ could be represented as matrices of
generators of rotations in the Lorentz group $\sf{SO(3,1)}$ so that
explicitly
\begin{equation}\label{a11}
\lambda^{(R)}_i{}^{\alpha\beta} = \epsilon_i{}^{\alpha\beta}\,,
\quad\lambda^{(R)}_i{}^{\alpha0} =0\,, \quad \alpha,\beta=1,2,3
\end{equation}
Substituting this in (\ref{a10}) one can easily check that there are
two solutions
\begin{equation}\label{a12}
h_j^{0\beta} =f(t) \, \delta_j^\beta\,, \quad h_j^{\alpha\beta} =
g(t)\, \epsilon_i{}^{\alpha\beta}
\end{equation}
which is the ansatz used in the main text.

\begin{acknowledgments}
We thank Sabine Hossenfelder for correspondence. The work of J.\
Kowalski-Glikman was supported in part by the grant
182/N-QGG/2008/0,  the work of R.\ Durka was supported by the
National Science Center grants  N202 112740 and 2011/01/N/ST2/00409
and by the European Human Capital Programme.
\end{acknowledgments}

\end{document}